\documentclass[useAMS,usenatbib]{mn2e}
\usepackage[dvips]{graphicx}
\usepackage{textcomp}
\usepackage{times}
\title[Possible spatial asymmetry in SRV UZ Arietis]{Possible spatial asymmetry in semi-regular variable UZ Arietis}
\author[Tapas Baug, T. Chandrasekhar and Shashikiran Ganesh]{Tapas Baug\thanks{E-mail: tapasb@prl.res.in (TB); chandra@prl.res.in (TC); shashi@prl.res.in (SG)}, T. Chandrasekhar\footnotemark[1] and Shashikiran Ganesh\footnotemark[1]\thanks{This file has been amended to highlight the proper use of \LaTeXe\ code with the class file.}\\Physical Research Laboratory, Ahmedabad 380009, India\\}
\begin{document}

\date{Submitted}

\pagerange{\pageref{firstpage}--\pageref{lastpage}} \pubyear{2014}

\maketitle

\label{firstpage}
\begin{abstract}
Semi-regular variables (SRVs) though closely related to Mira variables, are a less studied class of AGB stars. While asymmetry in the brightness distribution of many Mira variables is fairly well known, it is detected only in a few SRVs. Asymmetry in the brightness distribution at the level of a few milliarcsecond (mas) can be detected by high angular resolution techniques like lunar occultations (LO), long baseline interferometry, and aperture masking interferometry. Multi-epoch LO observations have the potential to detect a departure of brightness profile from spherical symmetry. Each LO event provides a uniform disk (UD) angular diameter along the position angle of the occultation. Any significant difference in the UD angular diameter values of multi-epoch LO observations signifies a brightness asymmetry. In this paper, we report for the first time three epoch UD angular diameter values of a SRV UZ Arietis using the LO technique at 2.2 $\mu m$. Optical linear polarization of the source observed by us recently is also reported. The asymmetric brightness distribution of UZ Ari suggested by a small difference in the fitted UD values for the three epochs, is discussed in the context of optical polarization exhibited by the source and the direction of polarization axis in the plane of the sky.
\end{abstract}


\section{Introduction}
Semi-regular variables (SRVs) are a sub-class of asymptotic giant branch (AGB) stars which also include long period variables called Mira variables. These objects show a large amplitude visual variability (V$_{amp} \ge$ 2.5) with well defined periods of 300-400 days. On the other hand SRVs have smaller variability amplitude (V$_{amp} <$ 2.5) and periods range from 10 to 1000 days. SRVs are also less studied than Mira variables. There is a speculation that SRVs could be the progenitors of Mira Variables, but needs confirmation (Bedding \& Zijlstra, 1998). The boundary between these two subclasses is not well defined. Several stars which were classified as Mira variable, have been reclassified as SRV (SRb) after observing irregularity in there periods (Ragland et al., 2006).

Asymmetry in AGB variables, particularly in Miras, have been observed earlier with high angular resolution techniques like lunar occultations (LO), long baseline interferometry (LBI), and aperture masking interferometry (Mondal \& Chandrasekhar, 2002; Ragland et al., 2006; Weiner et al., 2006). The departure from spherical/circular symmetry may be due to stellar rotation, non-radial pulsation or bright hot/cold spots in the outer layers of the atmosphere. Multi-epoch angular diameter measurements of the same source using LO technique are important in this context as they provide angular diameter values at different position angles across the occulted stellar source. Any significant variation in the angular diameter values is an indication of asymmetric brightness distribution in the occulted source. Such asymmetric brightness distributions have been reported earlier in several Miras. In a study using HST images, Karovska et al. (1997) reported asymmetry in Mira A atmosphere at 0.5 $\mu m$ which they ascribed to bright spots on the surface of the star, while the asymmetry observed in UV-band is explained as Mira A's interaction with Mira B. Asymmetric structure in a Mira variable, U Ori, has been seen using LO technique by Mondal and Chandrasekhar (2004). This spatial asymmetry was attributed to pulsational mass-loss in a preferred direction. Asymmetry in Mira variables often changes over their pulsation periods which is because of directional mass-loss in different pulsation periods.

In an extensive study, Ragland et al. (2006) observed 56 AGB variables using the three-telescope infrared optical telescope array (IOTA) interferometer at near infrared H-band (1.65 $\mu m$) to study asymmetry in the source brightness distribution. Their sample includes 35 Mira variables, 18 SRVs and 3 Irregular variables. Asymmetric brightness distribution has been detected in 29\% of the sources. They also speculated that 75\% of the Mira variables in their sample could be asymmetric, but are beyond the spatial resolution of the detector system they used. Among the 16 sources which show asymmetric brightness distributions, 3 are SRVs, namely, UU Aur, V Hya and Y CVn. All of the SRVs are carbon stars and have uniform disk (UD) angular diameters of 10.9 milliarcsecond (mas), 21.2 mas and 14.1 mas respectively. The cause of the asymmetry is not clearly known. They explained that it is unlikely for these stars to have rotational asymmetry or non-radial pulsation unless they have a relatively massive companion ($>$ 0.1 M$_\odot$). These stars are large in size and harbour massive envelopes around them. Hence, they cannot have sufficient angular momentum to develop asymmetry without a binary companion. Though convective cells can stimulate non-radial pulsations, but that will be on a smaller scale. However, they speculated that the asymmetry seen in Mira variables is due to the formation of an inhomogeneous translucent molecular screen located at about 1.5 to 2.5 stellar radii.

AGB variables generally develop molecular layers in the atmosphere which may cause different optical depth in different wavelengths. Wittkowski et al. (2011) performed spectro-interferometry on four Mira variables in the near-infrared (NIR) K-band and found wavelength dependent angular diameter. They also found a significant deviation from the point symmetry in H$_2$O band near 2.0 $\mu m$, which was interpreted as inhomogeneities or presence of clumps within water vapour.

Another aspect to be investigated while studying asymmetry in SRVs is the polarization of the source. Radiation from the luminous AGB variables is often found to be linearly polarized, generally 1-5\%. Asymmetric radiation field due to localized hot/cool spots or deformed shape of the envelope may provide an overall polarization. In case of AGB variables polarization arises from purely scattering phenomena. In a recent paper, Neilson, Ignace \& Henson (2013) has shown that Mira V-band polarization varies only 0-1\% over last 40 years but the variation in position angle corresponding to it is from 0$^\circ$ to 150$^\circ$. They suggest that these variations in  polarization as well as position angle of Mira are due to convective hot spots. On the otherhand in a SRV (V CVn), they found variability in the polarization of 0-7\% with a nearly constant position angle. The constant position angle of polarization suggests a long term asymmetry along the direction of polarization.

Polarization in very cool stars have been studied for a longtime (Serkowski, 1966; Dyck, Forbes \& Shawl, 1971; Kruszewski \& Coyne, 1976). Kruszewski \& Coyne (1976) reported polarization measurement of 18 cool stars (including UZ Ari) in the visible and NIR bands in the spectral range of 0.3-1.0 $\mu m$. They made polarization observations with the Steward observatory 229 cm reflector on the Kitt Peak and the 154 cm reflector of the Mt. Lemmon observatory. They found a good correlation of degree of polarization with [3.5 - 11 $\mu m$] colour which characterizes the dust envelope of the star. On the otherhand, a weak correlation is found for polarization with NIR (I-K) colour which characterizes the stellar photosphere. In addition to polarization, maser emission also is not an uncommon phenomena among the late type giants. Circular polarization observations of these masers have revealed the strength and structure of the magnetic field through out the circumstellar shell. Kruszewski \& Coyne (1976) found a good correlations between the degree of polarization and the separation between 18 cm OH-maser line components. The authors concluded that the polarization is produced in the extended stellar atmosphere where the infrared excess and OH-maser emission also originate. However, the source of our interest UZ Ari, does not have any OH-maser emission detection till date, though water maser at 22 GHz has been seen sporadically (Engels, Schmid-Burgk \& Walmsley, 1988; Takaba et al., 2001).

In this paper we present 3-epoch uniform disk angular diameter measurements of the semi-regular variable UZ Arietis using the LO technique in the NIR K-band (2.2$\mu m$). These measurements are the first reported high angular resolution observations of UZ Ari. We have made a detailed study of the SRV UZ Arietis based on our LO observations and optical polarimetric observations made at a latter epoch.

\begin{table*}
\centering
\begin{minipage}{200mm}
\caption{Details of source and the three observed LO events of UZ Ari}
\begin{tabular}{@{}clcccccccc@{}}
\hline
\multicolumn{2}{l}{Periodicity}   & \multicolumn{2}{l}{158$\pm$10 days}    & ASAS$^\dag$  \\
\multicolumn{2}{l}{Magnitudes}    & \multicolumn{2}{l}{V 11.6 - 12.7}      & ,,           \\
                                & & \multicolumn{2}{l}{V 11.2 - 11.8}      & L1974$^\dag$ \\
\multicolumn{2}{l}{          }    & \multicolumn{2}{l}{J 2.87$\pm$0.06}    &\multicolumn{4}{l}{This paper (Observed on JD 245 6555.3)}\\
\multicolumn{2}{l}{          }    & \multicolumn{2}{l}{H 1.76$\pm$0.08}    &   ,,\\
\multicolumn{2}{l}{          }    & \multicolumn{2}{l}{K 1.22$\pm$0.05}    &   ,,\\
\multicolumn{2}{l}{Spectral type} & \multicolumn{2}{l}{M7.8 - M8.1}        & ,,           \\
\multicolumn{2}{l}{Distances}     & \multicolumn{2}{l}{$\sim$140 pc}       & C1995$^\dag$ \\
\hline
\hline
   JD Obs. & Detector used   &Lunar Phase$^a$& Altitude &  PA$^b$  & V$_{comp}^c$ & Sampling  &     K-mag     &Sky Conditions\\
  2450000+ &                 &    (days)     &($^\circ$)&($^\circ$)& (m ms$^{-1}$)& time (ms) &               &              \\
\hline
 4100.2108 & InSb Photometer &      10.1     &   69.5   &    63    &     0.67     &   2.000   & 1.37$\pm$0.05 & Photometric  \\
 5166.3799 & HgCdTe Sub-array&      14.1     &   40.8   &    98    &     0.78     &   8.926   & 1.49$\pm$0.30 & Spectroscopic\\
 5221.4414 & HgCdTe Sub-array&       9.6     &   46.4   &   123    &     0.56     &   8.926   & 1.14$\pm$0.30 & Photometric  \\
\hline 
\end{tabular}
\begin{flushleft}
$^a$ Lunar Phase measured in days from the new moon.\\
$^b$ PA is the position angle of the point of occultation on the lunar limb measured from North through East (NESW).\\
$^c$ V$_{comp}$ refers to the predicted velocity component of the moon in the direction of occultation.\\
$^\dag$ASAS: ASAS Catalog of variable stars; L1974: Lockwood, 1974; C1995: Celis, 1995\\
\end{flushleft}
\label{obs_detail}
\end{minipage}
\end{table*}

\section[]{Observations and Data Analysis}
\subsection[]{Source Details and Earlier observations}
The source, UZ Ari is a semi-regular variable with a period of 163 days and the V-band magnitude varies from 11.80 (V$_{max}$) to 12.60 (V$_{min}$) according to GCVS catalog (Samus et al., 2009). The source is classified as M8 giant and has a circumstellar shell which shows broad oxygen-rich dust emission in IRAS low resolution spectra (Sloan \& Price, 1998). Lockwood (1974) reported V$_{max}$ of 11.2 and V$_{amp}$ of 0.6 mag, and a corresponding change in spectral type from M7.8 to M8.1, but the observations were not complete enough to report any definitive period. Later Celis (1995) reported maximum and minimum V-magnitudes of 11.2 and 12.2 for the source. A V-band light curve spanning about seven years has been reported in ASAS catalog of variable star (Pojmanski, Pilecki \& Szczygeil, 2005). Though there is no report of the periodicity in the catalog, we calculated a period of 158$\pm$10 days from the light curve. It is in good agreement with the periodicity of 163 days, reported in GCVS catalog. The details of the source are listed in Table.\ref{obs_detail}.

Water maser (22 GHz) has been detected on this source sporadically. First H$_2$O maser was detected on this source by Crocker and Hagen (1983) using 36.6 m antenna of Haystack Observatory. They found an H$_2$O maser flux of 4.2 $Jy$ on March 1980. Engels, Schmid-Burgk \& Walmsley (1988) also reported H$_2$O maser flux of 5.6 $Jy$ using 100 m radio telescope at Effelsberg on April 1984. An upper limit of H$_2$O maser of $<$7.7$\pm$2.6 $Jy$, observed on April 1988 has also been reported by Comoretto et al. (1990). Later Takaba et al. (2001) detected H$_2$O maser again and reported a flux of 2.5 $Jy$ observed on November 1991. However, there is no report of detection of other maser species like SiO or OH-masers on UZ Ari in the literature. Though water maser has been detected many times earlier on the source, surprisingly Kim, Cho \& Kim (2013) reported no detection of H$_2$O maser and provided an upper limit of 1.1 $Jy$, observed in 2009 November.

Estimation of linear radius of a star from angular diameter measurements requires a good distance value. The source has no parallax measurement because it is fainter (m$_V$ $\sim$ 11.2) than the Hipparcos parallax measurement limit (m$_V$ $\sim$ 9). Two indirect distance estimations are available in the literature. Iyenger et al. (1982) reported an upper limit of the distance to the source of 200 pc using magnitude modulus. Later Celis (1995) estimated the distance to the source based on analytical relationship between the absolute magnitude and the spectral type. The author calculated the distance to UZ Ari to be 140 pc using spectral type of M8.1. The spectral type was obtained from their observed spectra.

The source exhibits a substantial amount of optical linear polarization ($\sim$3\%) in V-band, increasing towards shorter wavelengths (Dyck, Forbes \& Shawl, 1971; Kruszewski \& Coyne, 1976, and this work). The observational details are discussed in later sections.

\subsection[]{Multi-epoch Lunar Occultation Observations}
All the LO observations of this source were carried out in the NIR broad-$K$-band (2.2 $\mu m$/ band-width 0.4 $\mu m$) using 1.2 m Mt. Abu telescope. The details of the LO observations are listed in Table.\ref{obs_detail}. All are disappearance events at lunar phase measured in days after new moon also listed in the table. One event (2006 December 30) was observed using a single element InSb detector. The details of this detector system can be found in Chandrasekhar (2005). The other two events (2009 November 30 and 2010 January 24) were recorded using the fast subarray mode of the NICMOS IR-camera.

The observed light curves are carefully modeled to extract the uniform disk (UD) angular diameter of the stellar sources using the method of non-linear least squares (NLS). NLS involves five free parameters, namely, geometric time of the occultation, velocity component of the moon in the direction of occultation, signal level, background level and the UD angular diameter. A second order polynomial is also used if needed to model the varying lunar back-ground  of the light curve. A $\chi^2$ minimization is followed to obtain the best estimation of the five parameters. The point source Fresnel diffraction pattern modulated by the finite telescope aperture, finite optical and electrical bandwidths of the system along with UD angular diameter are used to model the observed light curve.

\begin{figure}
\centering
\includegraphics[width= 8 cm, height =6 cm]{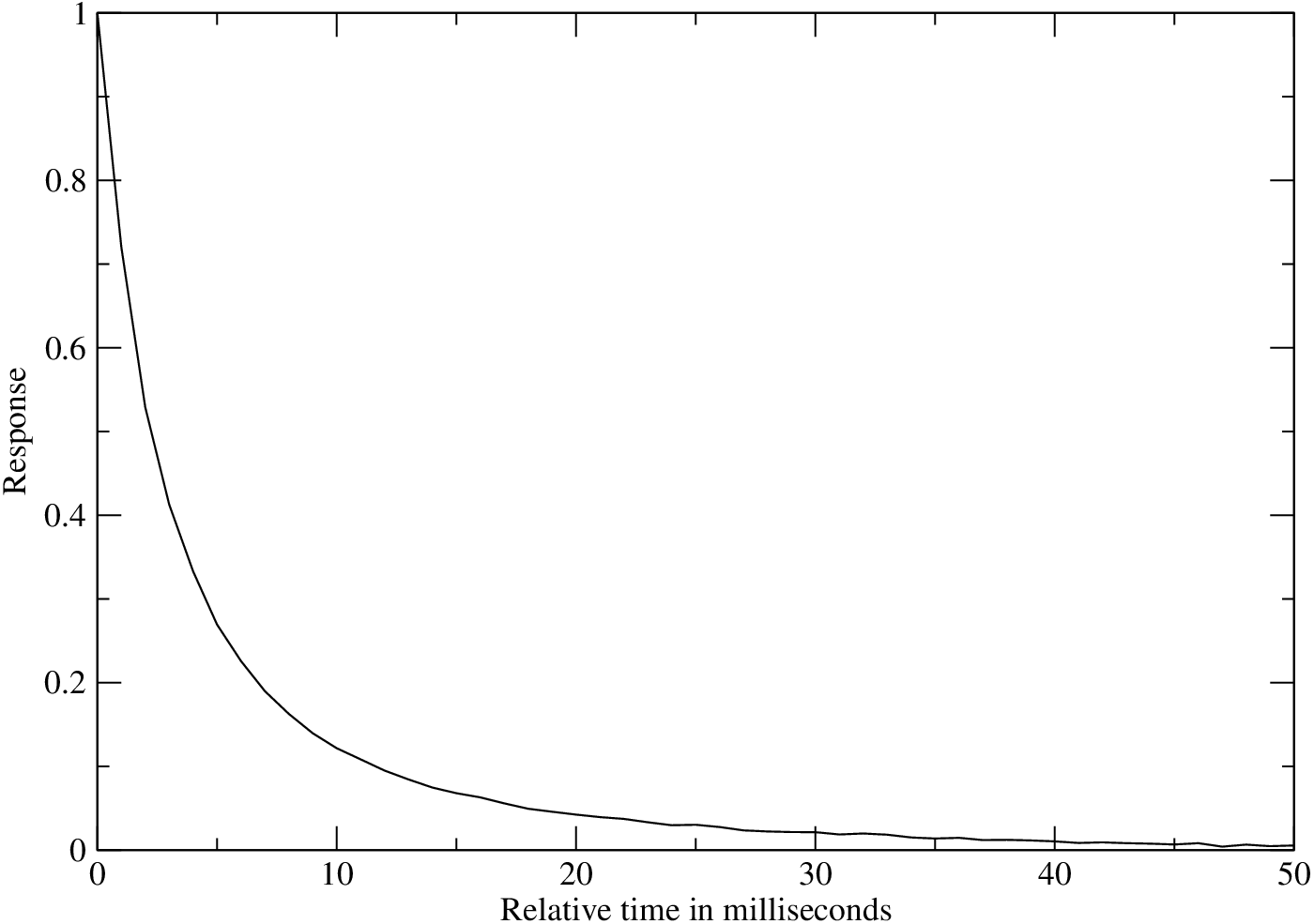}
\caption{Experimentally determined time response curve of the fast photometer.}
\label{timeres}
\end{figure}

While telescope aperture and filter band width do not change in the system, the electrical bandwidth which is determined by the time response of the photometer system changes with bias voltage given to offset sky background in the InSb detector and has to be determined exactly for every event. A fast pulsing light emitting diode (LED) which has better than microsecond response times is flashed into the photometer system and the square pulses recorded in the K-band. From the shape of the recorded pulse in the fast photometer the time response curve is derived for each event and included in the analysis. Fig.\ref{timeres} shows the time response curve of the photometer used in event 1. Apart from time response, scintillation noise in the photometer diaphragm, sky conditions, phase of lunation, lunar slope, altitude of observation, residual tracking errors of the telescope system can all contribute to the errors affecting the fitting of the model to the observed occultation light curve.

The sub-array mode which involves reading 10$\times$10 pixels, as rapidly as possible, requires 3 millisec  for integration, 3 millisec for reset and nearly 3 millisec electronic dead time leading to a data sampling time during occultation of 8.926 millisec. Inspite of our best efforts it was not possible to lower this sub-array sampling time for the NICMOS IR camera which remains the major draw back in using the IR camera for LO events. Details of the subarray mode of operation for lunar occultation observations and analysis of the light curves are extensively discussed in Chandrasekhar \& Baug (2010).

\begin{figure}
\centering
\includegraphics[width=8 cm,height= 6.0 cm]{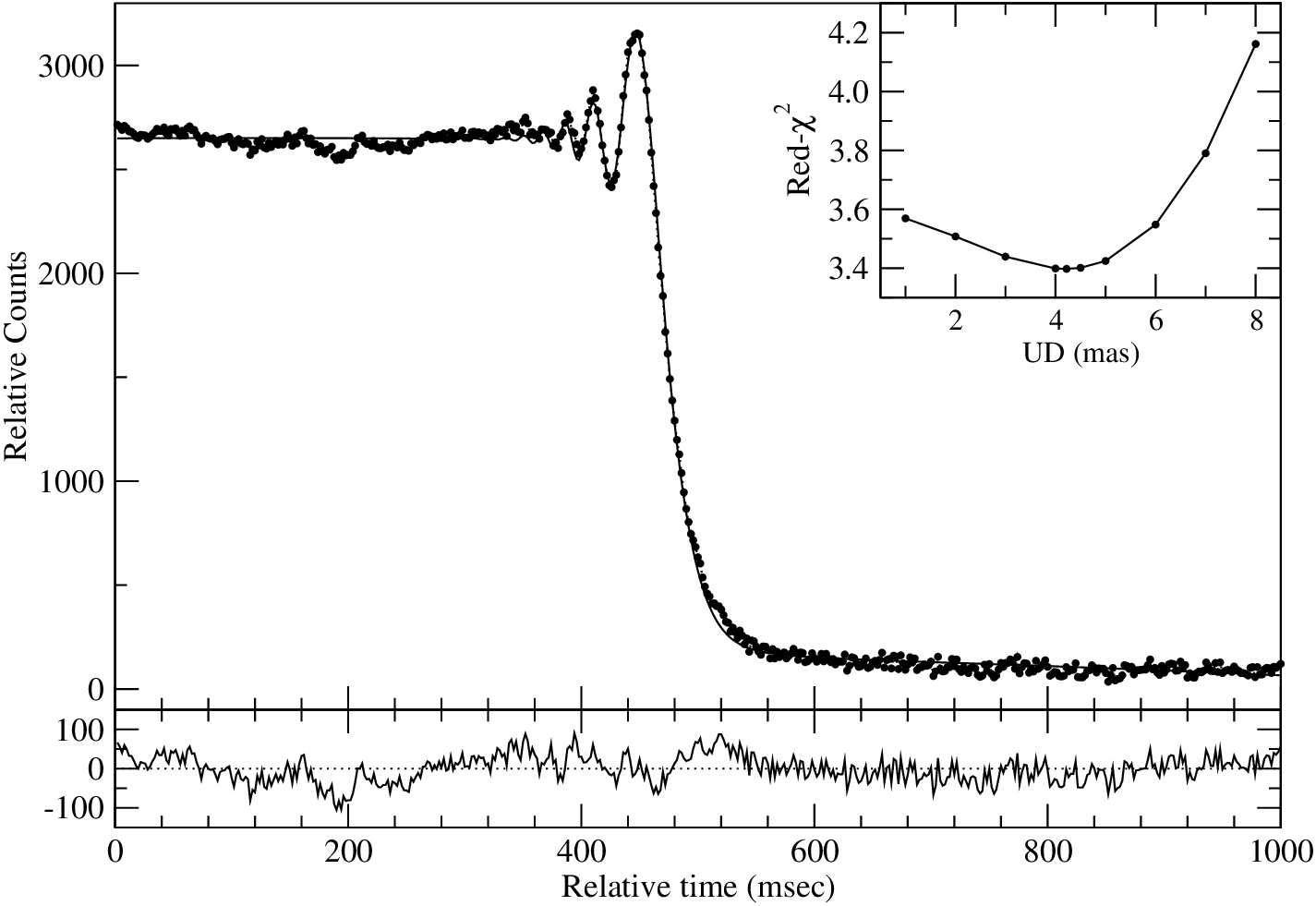}
\caption{The model fit to the light curve, observed on 2006 December 30. It fits to the UD angular diameter of 4.5$\pm$0.5 mas. The lower panel shows the residuals (data\textminus model), while the inset shows the error curve. The error curve is flatter towards lower angular sizes as the UD value is close to the resolution limit of about 2.5 mas. It rises sharply beyond 5.0 mas.}
\label{lc1}
\end{figure}

\begin{figure}
\centering
\includegraphics[width=8 cm,height= 6.0 cm]{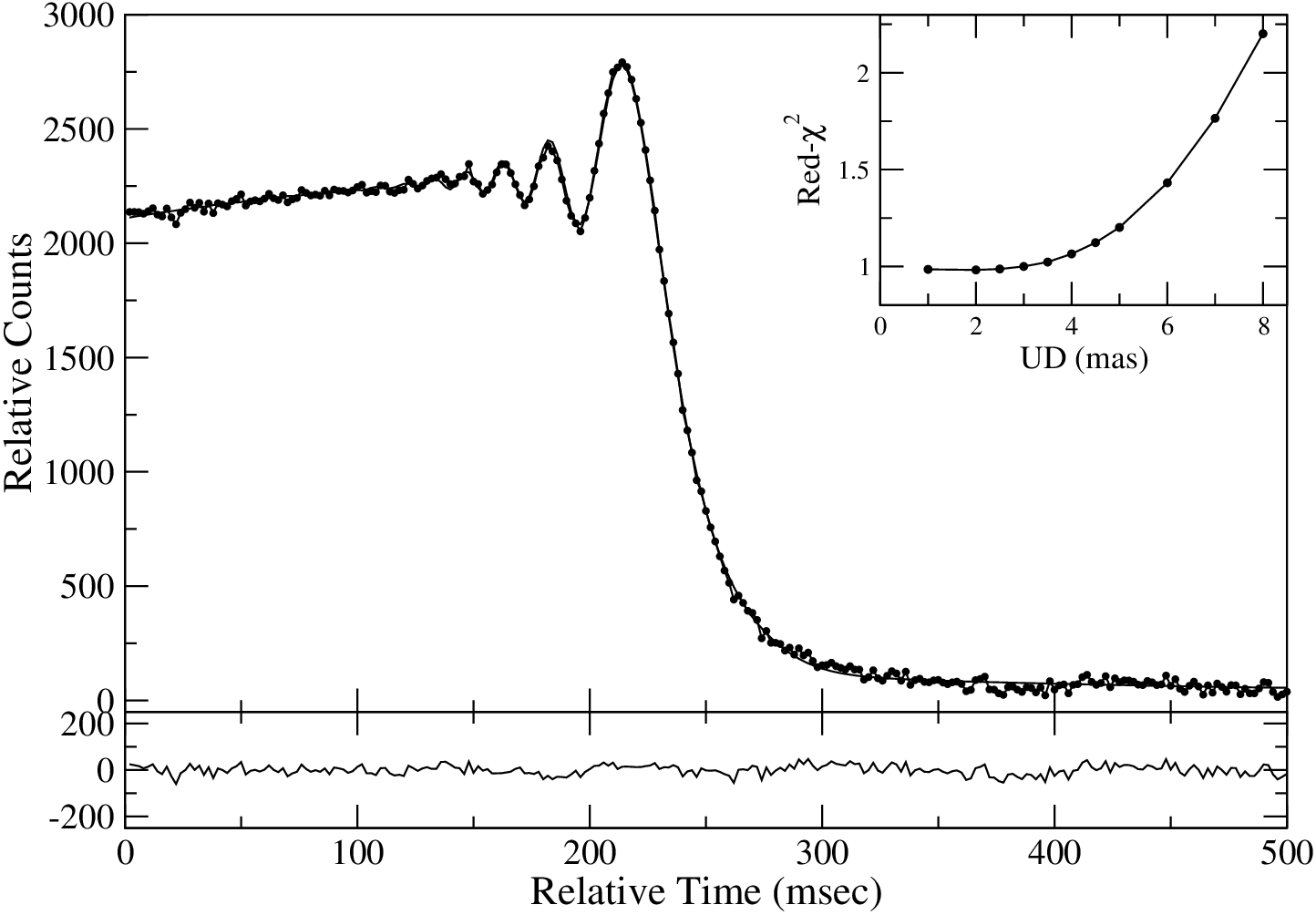}
\caption{LO light curve of bright star $\alpha$ Leo which is an unresolved point source for the photometer. The limit of resolution is 2.5 mas. Note the flat portion of error curve below 2.5 mas showing unresolved nature of source.}
\label{lc2}
\end{figure}

\begin{figure}
\centering
\includegraphics[width=8 cm,height= 6.0 cm]{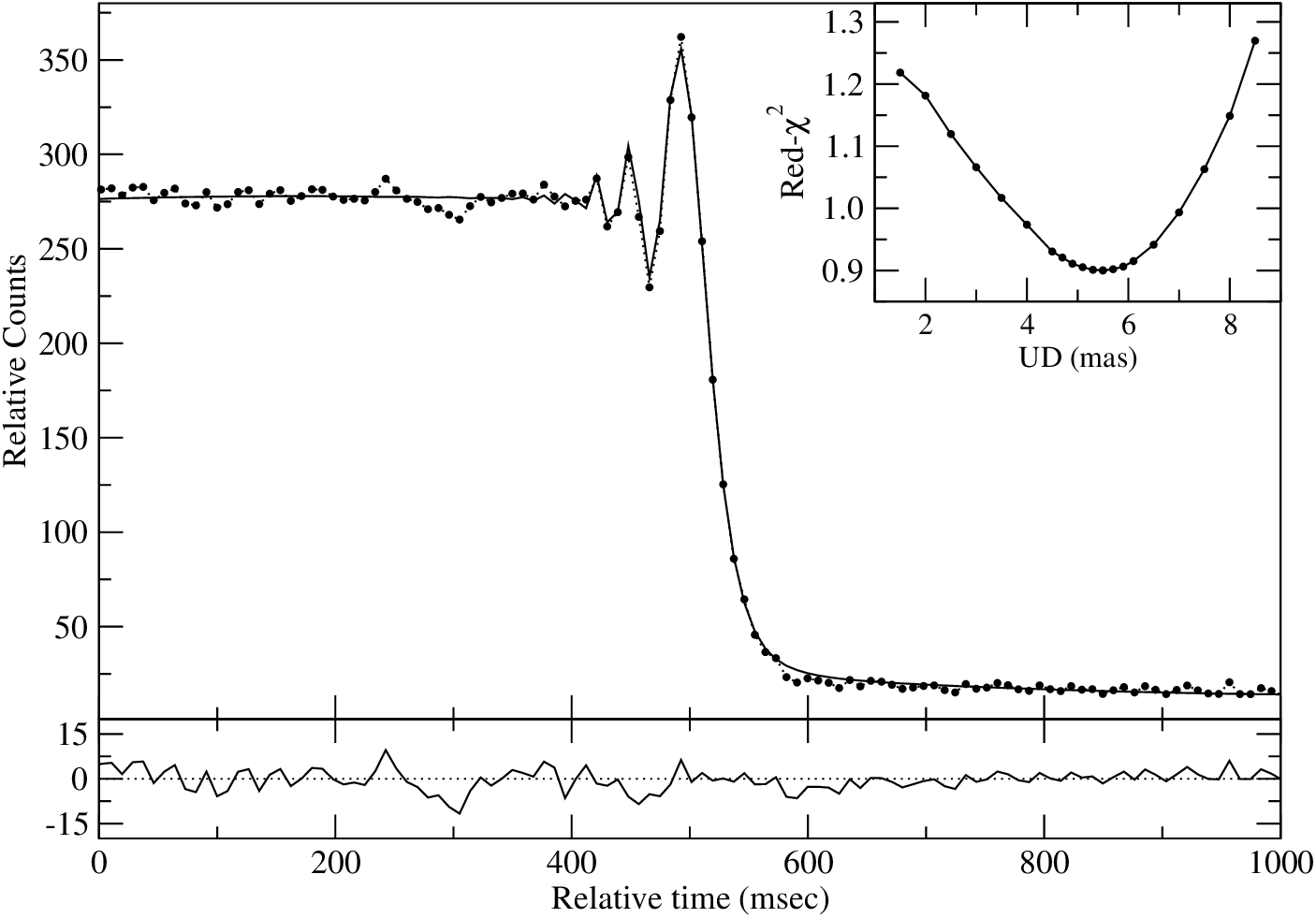}
\caption{The model fit to the light curve, observed on 2009 November 30. It fits to the UD angular diameter of 5.5$\pm$0.5 mas. The lower panel shows the residuals (data\textminus model) and the inset shows error curve. As the UD angular size is well above the resolution limit the error curve is almost symmetrical.}
\label{lc3}
\end{figure}

\begin{figure}
\centering
\includegraphics[width=8 cm,height= 6.0 cm]{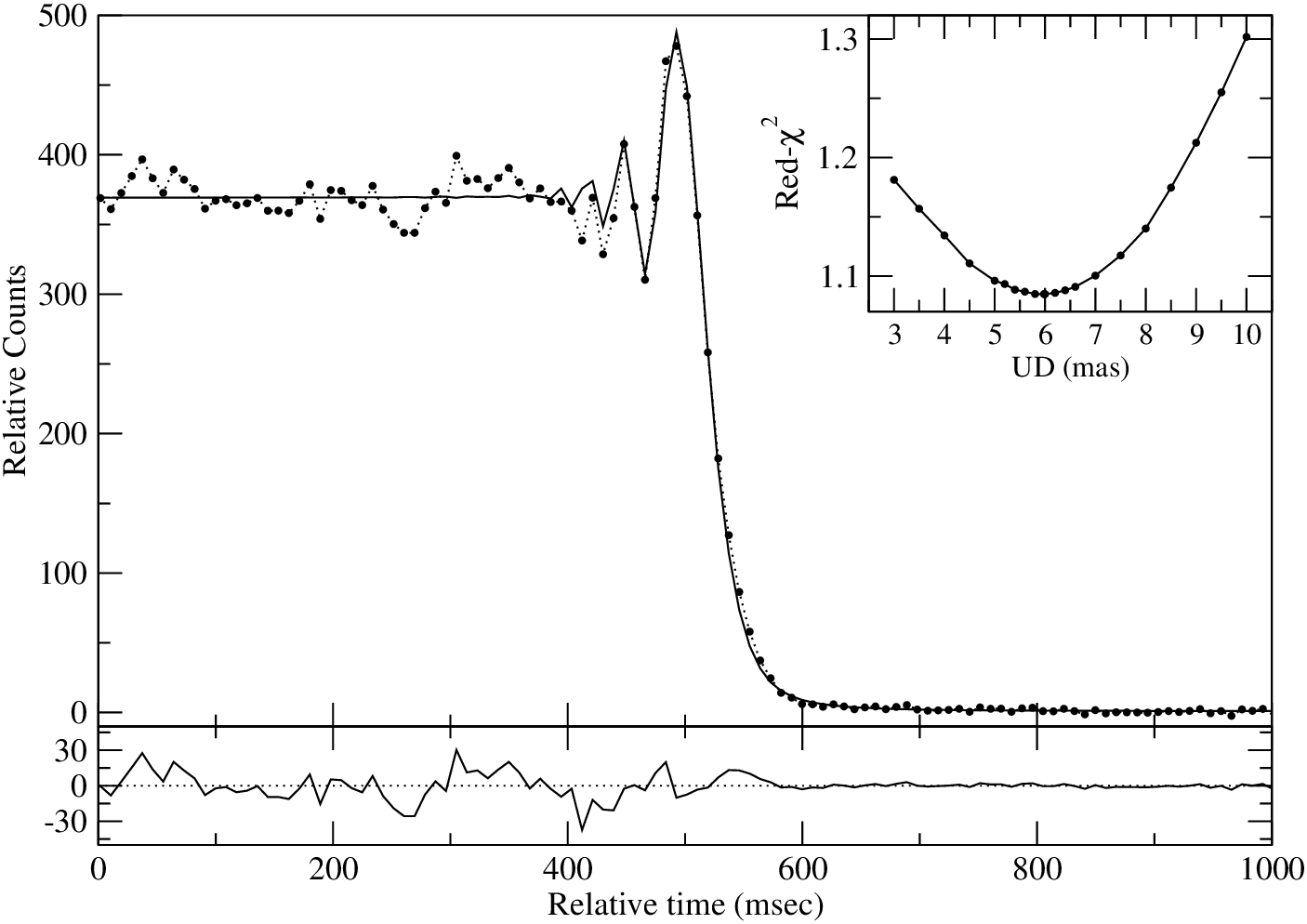}
\caption{The model fit to the light curve, observed on 2010 January 24. It fits to the UD angular diameter of 6.0$\pm$0.5 mas. The lower panel shows the residuals (data\textminus model) and the inset shows error curve. As the UD angular size is well above the resolution limit the error curve is almost symmetrical.}
\label{lc4}
\end{figure}

The observed and corresponding model fitted light curves are shown in Figures \ref{lc1}, \ref{lc3} and \ref{lc4} for 2006, 2009 and 2010 events respectively. The residuals (data\textminus model) are also plotted in the respective lower panels. The red-$\chi^2$ values with respect to different UD angular diameter values are also shown in the inset of each figure. The skewed nature (flattening towards smaller angular sizes) of the error curve for epoch 1 despite the high signal to noise ratio (S/N) of the observations is because in this case we are approaching the resolution limit of our LO observations which is 2.5 mas. To illustrate this aspect LO light curve of a bright point source $\alpha$ Leo is also included as Fig.\ref{lc2}. It can be seen that the error curve for this star flattens out below 2.5 mas which is our resolution limit. It may be noted that the signal levels are actual observed counts and not normalised to a uniform value for the three events. Hence, the best fitted model curves for the three events correspond to their lowest red-$\chi^2$ values which are different for each event. The UD angular diameter values obtained from three LO events are listed in Table.\ref{results}. The best fit UD value of UZ Ari for epoch 1 of 4.5 mas has actually less error than 0.5 mas towards higher angular sizes due to the sharply rising error curve. However, we have conservatively retained the same error value of 0.5 mas applicable towards smaller UD values so that the UD value of epoch 1 reads 4.5$\pm$0.5 mas. The error curves of the other two epochs sampled at coarser time intervals do not exhibit the skewed nature as they are both well above the resolution limit. The UD values of UZ Ari for epoch 2 and epoch 3 are respectively 5.5$\pm$0.5 mas and 6.0$\pm$0.5 mas.

\subsection[]{Polarization measurements}
Optical polarization observations have also been performed recently on this source using the photo-polarimeter (PRLPOL) mounted on the Cassegrain focus of the 1.2 m Mt. Abu telescope. This instrument works on the rapid modulation principle. The details of the polarimeter can be found in Deshpande et al. (1985). Ganesh et al. (2009) describe recent upgrades to the system. The instrument has an 8 position filter slide which normally contains standard UBVRI and a few narrow band filters. The observations were carried out on 2013 October 16 and 2014 January 06. Polarization observations in both the epochs were carried out in 4-filter bands (UBVR). A bright moon was present during the observation of 2013 October 16 and due to moonlit conditions only V- and R-band observations turn out to be usable. BVR polarization observations taken on 2014 January 06 (6 days after new moon) are usable. U-band magnitude of the source was comparable to the sky brightness on those nights. During both runs, the observing sequence of Sky-Object-Sky and Sky-Object-Object-Object-Sky was followed. Thus sky corrections were made using nearby background sky before and after the object observations. Details of the polarization measurements are listed in Table.\ref{pol_data}.

Two earlier polarimetric measurements are reported on this source by Dyck, Forbes \& Shawl (1971) and Kruszewski \& Coyne (1976). Dyck, Forbes \& Shawl (1971) measured the polarization in the visible and NIR bands. The source showed a monotonic rise of polarization towards the shorter wavelengths. But it has not shown any variation in the position angle ($\sim$ 130$^\circ$) with wavelengths, like in the case of another SRV, V CVn reported by Neilson, Ignace \& Henson (2013). The source, UZ Ari, is nearby ($\sim$140 pc) and situated at high Galactic latitude (\textminus31$^\circ$). So, the observed polarization is unlikely to be interstellar but intrinsic in nature. After a few years, Kruszewski \& Coyne (1976) also measured polarization in seven bands in the wavelength region of 0.3 $\mu m$ to 1.0 $\mu m$. Though they obtained a smaller polarization than it was measured by Dyck, Forbes \& Shawl (1971), but the position angle and the shape of the wavelength dependence of polarization are the same in both sets of data. They also have reported that the polarization is intrinsic and is the largest in the ultra-violet.

\subsection[]{Near-IR Photometry and Spectroscopy}
Photometric and spectroscopic observations in the NIR J (1.25 $\mu m$), H (1.65 $\mu m$) and K-bands (2.20 $\mu m$) have been carried out on this source using the NICMOS IR Imager/Spectrometer. Aperture photometry observations were carried out recently (JD 245 6555.3) by dithering the source in five positions. Accurate JHK magnitudes obtained by us for this epoch are J$=$2.87$\pm$0.06, H$=$1.76$\pm$0.08, and K$=$1.22$\pm$0.05. These values are also listed in Table.\ref{obs_detail}. We also estimated the NIR K-band magnitude of the source on the day of occultation from the observed LO light curves and they are listed in Table.\ref{obs_detail}. For epoch 1 we get a good value of 1.37$\pm$0.05 while the errors are larger for the others 1.49$\pm$0.30 (epoch 2) and 1.14$\pm$0.30 (epoch 3).

Spectroscopic observations were recorded on three epochs, 2010 January 28 and 29, and 2013 September 19 and analysed using {\sc iraf} package. The average velocity resolution in NIR JHK bands is about 300 km s$^{-1}$. The JHK spectra are typical of a M8 giant and do not exhibit any significant departures from spectra of evolved giants.

\begin{figure}
\centering
\includegraphics[width=8 cm,height= 6 cm]{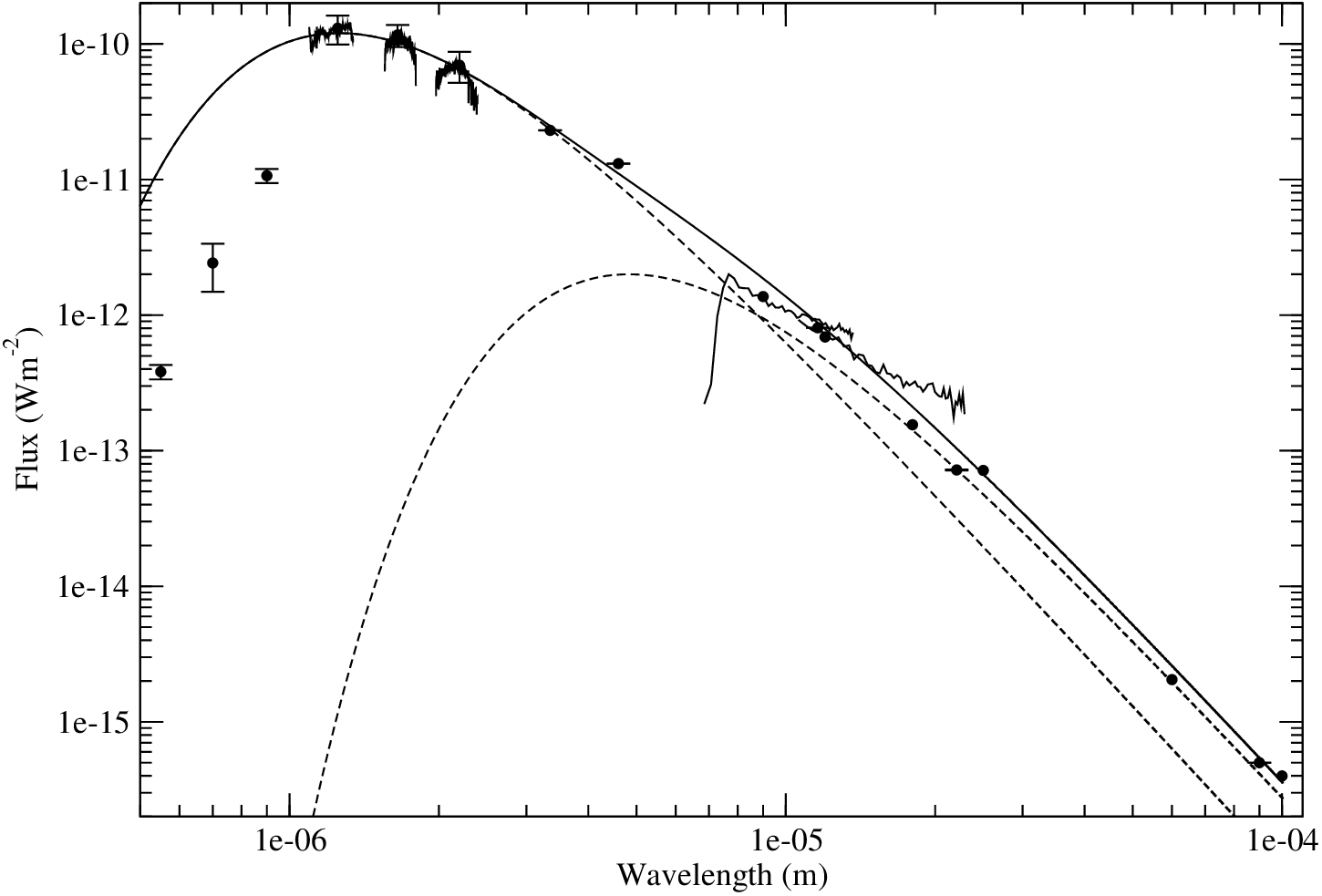}
\caption{The spectral energy distribution of the source. A simple two temperature black-body (2300 K and 600 K) has been fitted to the data. Our observed NIR J, H, K-band spectra and IRAS LRS are also overplotted.}
\label{sed}
\end{figure}

A simple two temperature black-body fit has been performed on spectral energy distribution curve of the source constructed from flux values available in the literature. We also overplotted our observed spectra and IRAS low resolution spectra (IRAS Science Team, 1986) on it. The stellar temperature has been assumed to be 2300 K to be consistent with its spectral type of $M8$ $III$. A shell temperature of 600 K is added to fit the data points beyond the K-band (Fig.\ref{sed}). It is estimated that about 1\% of the total flux in the NIR K-band is contributed from the shell. The contribution of the shell increases beyond the K-band and is about 40\% of the total flux value at 10 $\mu$m. The source exhibits [3.35 \textminus 11.6$\mu m$] colour excess of 1.93$\pm$0.05. This colour excess  along with other Far IR observations indicates the presence of circumstellar material though it is not possible to conclude from the spectral energy distribution alone whether it is a shell or a torus or a disk.

Some other interesting aspects of this sources are listed below. It has faded almost in all wavelength bands by 0.5 magnitude in last 30-40 years. Though water maser has been detected on this source sporadically, but in a recent paper, Kim, Cho \& Kim (2013) could not detect water maser emission even with lower detection limit than earlier. NIR (J-K) colour for 2MASS, DIRBE and our observations are 1.55, 1.51 and 1.64 respectively. Also no significant change in colour is found when we examined IRAS [12 \textminus 25] and WISE [11.6\textminus 22.1] colour and the corresponding values are 0.69 and 0.87 respectively. But mid-IR [3.5\textminus 11.0 $\mu m$] colour appears to show a variation. Dyck, Forbes \& Shawl (1971) reported [3.4\textminus 10.2] is 1.23, and Kruszewski \& Coyne (1976) reported [3.5\textminus 11.0] colour 1.3. However, the WISE catalog reports [3.35\textminus 11.6] colour of 2.62. The deviation in the mid-IR colour may be because of different effective wavelength or different effective bandwidth. Earlier Iyenger et al. (1982) also found an huge excess emission in 19.8 $\mu m$ without any corresponding excess in 11.0 $\mu m$. The reason for this unusual behaviour of the source is unknown.

\begin{table*}
\centering
\caption{Fitted Uniform disk K-band angular diameters from  observed LO light curves.}
\begin{tabular}{@{}lccccccc@{}}
\hline
 JD Obs.      & Photometric & S/N &    PA    & N$_{data}$&  Obs. UD   & Linear  \\
245 0000+     & Phase       &     &($^\circ$)&           &  Dia (mas) & Radius (R$_{\odot}$)  \\
\hline
 4100.2108    &0.38$\pm$0.06& 118 &    63    &   500     & 4.5$\pm$0.5& 68$\pm$8 \\
 5166.3799    &0.13$\pm$0.06&  65 &    98    &   112     & 5.5$\pm$0.5& 82$\pm$8 \\
 5221.4414    &0.48$\pm$0.06&  39 &   123    &   112     & 6.0$\pm$0.5& 90$\pm$8 \\
\hline 
\end{tabular}
\begin{flushleft}
\end{flushleft}
\label{results}
\end{table*}

\section{Results and Discussion}
The derived uniform disk angular diameters for the three epochs along with the other details like Signal to Noise ratio, number of data points used in the analysis, position angle of the source for each epoch are given in Table.\ref{results}. Also listed in the table are the linear radii of UZ Ari for each epoch adopting a distance of 140 pc as discussed in Section 2.1. The light curve observed on 2006 December 30 using the fast InSb photometer, fits to an angular diameter value of 4.5$\pm$0.5 mas. The other two light curves obtained using NICMOS sub-array fit to UD values of 5.5$\pm$0.5 mas and 6.0$\pm$0.5 mas respectively. Even though the differences in the fitted UD values for the three epochs are small but it may be noted from error curves of Figures 4 and 5 that neither epoch 2 nor epoch 3 observations can be fitted to values below 5.0 mas. Clearly epoch 1 observations yield a smaller uniform disk angular diameter compared to epoch 2 or 3. The difference could be a real change in angular size due to pulsation of the atmosphere as in the case of a Mira like $\chi$ Cyg (Lacour et al., 2009) or it may be due to position angle dependent geometrical effect brought about by the presence of a disk or torus extending in a particular direction. In case of pulsational effect the derived UD values should vary with photometric phase of the star. We consider both possibilities.

The photometric phase of each epoch has been derived using the periodicity of 158$\pm$10 days obtained from the ASAS V-band light curve (Pojmanski, Pilecki \& Szczygeil, 2005) as discussed in Section 2.1. It is known that some Mira show variability in their angular diameter with photometric phase. Cycle-to-cycle variability of angular diameter has been reported for a few Mira variables by Woodruff et al. (2008).  Lacour et al. (2009) found about 30\% intra-cycle variation in the NIR H-band angular diameter of a Mira variable, $\chi$ Cyg which has a period of 408 days and shows a large V-band variability of nearly 11 magnitudes. In contrast the V-band amplitude of UZ Ari is only about one magnitude and its K-band variation is only about 0.15 magnitude. It is not established that semi-regular variables like UZ Ari with a small amplitude variation in their light curve would show a phase dependent variation of angular diameter like the Miras.

In case of phase dependent variation, the angular diameter is anti-correlated with the optical phase. The calculated  phases of UZ Ari during the three epochs are 0.38, 0.13 and 0.48 respectively. It can be seen that epoch 1 and 3 are closer to the minima (phase 0.5), while epoch 2 is near maximum phase. So, if angular diameter corresponds to the photometric phase, we expect to see epoch 1 angular diameter to be larger than epoch 2 values. For epoch 3 the phase is close the minima but the UD values are not different from epoch 2 values within errors of observation. Hence we are led to the conclusion that no correspondence of angular diameter with photometric phase has been observed for UZ Ari from our LO observations. The variation of UD  angular diameter of UZ Ari is not phase dependent and is due to other causes connected with existence of significant polarization. We argue that the UD angular size variation of UZ Ari is position angle dependent because the star has a circumstellar material in the form of a disk or torus along an axis of polarization close to position angle of 123$^\circ$.  

We next take up the discussion of the polarisation measurements. Recent high quality aperture-masked polarimetric interferometry of a few AGB stars have revealed dust shells at very small stellar radii (2 stellar radii) composed of large grains of about 0.3$\mu$m in radius. The grains are contributed by stellar pulsations (Norris et al., 2012) and the polarization observed is due to scattering of starlight by these grains. Hence, it is not unlikely that we are seeing from our limited LO measurements evidence of some type of circumstellar material with an alignment axis defined by the direction of polarization.

The R-band polarization values of UZ Ari available in the literature as well as our measurements are listed in Table.\ref{pol_old} for comparison. It is clear that the position angle of the polarization has not changed in last 35 years and the polarization value also has not changed significantly. A similar effect has been seen by Neilson, Ignace \& Henson (2013) for the SRV V CVn.

Long ago Kruszewski \& Coyne (1976) had reported a good correlation between the degree of polarization and [3.5\textminus 11 $\mu m$] colour which characterizes the dust envelope. UZ Ari has [3.5\textminus 11 $\mu m$] colour of 2.6 mag (WISE catalog). On the basis of this correlation the source is expected to show a substantial amount of intrinsic polarization in excess of 2\%. The polarization is caused by the scattering by dust grains present in the circumstellar envelope leading to an asymmetric brightness distribution, detected in our 3-epoch LO observations.

In the case of an asymmetric and inhomogenous disk or envelope around the star, the direction of polarization position angle would depend on the number density and size distribution of scattering particles. In the case of an optically thin shell, the direction of polarization position angle would be perpendicular to the plane of scattering or as seen by observer perpendicular to the direction of the largest dimension of the envelope. This is because single scattering would dominate.  In the case of an envelope or disk dominated by large fluffy grains, multiple scattering would happen and that would give rise to polarization with position angle parallel to the plane of scattering. While the source exhibits an IR excess (Fig.\ref{sed}) it is difficult to quantify the amount of circumstellar matter and conclude on whether it is optically thick or thin. However, the observed position angle of polarisation is close to the position angle for epoch 2 and 3, suggesting an axis of asymmetry along this direction.
 \begin{table}
 \centering
  \caption{Optical polarimetric measurements of UZ Ari using PRLPOL attached to 1.2 m Mt. Abu telescope.}
  \begin{tabular}{@{}lcccll@{}}  \hline
 Filter & JD Obs.  & Linear        & PA of        \\
 Band   &245 0000+ & Pol. (\%)     &Pol ($^\circ$)\\
\hline
   B    & 6664.31  & 1.99$\pm$0.90 & 132$\pm$12   \\
        &          &               &              \\
   V    & 6582.33  & 3.48$\pm$0.44 & 129$\pm$3    \\
        & 6664.25  & 3.29$\pm$0.53 & 128$\pm$4    \\
        &          &               &              \\
   R    & 6582.35  & 3.14$\pm$0.17 & 130$\pm$1    \\
        & 6664.23  & 3.32$\pm$0.22 & 132$\pm$2    \\
\hline 
\end{tabular}
\label{pol_data}
\end{table}

\begin{table}
 \centering
  \caption{R-band polarimetric measurements of UZ Ari}
  \begin{tabular}{@{}rcccll@{}}
  \hline
 JD Obs.  & Linear        & PA of        & References$^\dag$ \\
244 0000+ & Pol. (\%)     &Pol ($^\circ$)&            \\
\hline
  925.5   & 3.26$\pm$0.21 & 126          & D1970      \\
  951.5   & 3.43$\pm$0.18 & 130          & D1970      \\
 1643.8   & 2.30$\pm$0.06 & 130          & K1976      \\
16582.4   & 3.14$\pm$0.17 & 130$\pm$1    & This paper \\
16664.2   & 3.32$\pm$0.22 & 132$\pm$2    & ,,         \\
\hline 
\end{tabular}
\begin{flushleft}
$^\dag$D1970: Dyck, Forbes \& Shawl, 1970; K1976: Kruszewski and Coyne, 1976
\end{flushleft}
\label{pol_old}
\end{table}

\section{Conclusions}
From a three epoch LO observations at 2.2 $\mu$m we report the first uniform disk angular diameter measurements of semi-regular variable UZ Arietis. Three well resolved uniform disk angular diameters are obtained at three different position angles across the source. Optical linear polarization measurements made recently by us combined with earlier values suggest that source polarization value and direction has not changed over the last 40 years. Our LO measurements along with the  polarization observations suggest a small asymmetry in the brightness profile of the source directed along or close to the polarisation axis.
 
\section*{Acknowledgments}
This work was supported by Dept of Space, Govt of India. This research made use of the SIMBAD data base operated at the CDS, Starsbourg, France and catalogues associated with it,AAVSO and GCVS for the optical light curve, data products from the Wide-field Infrared Survey Explorer, a joint project of the University of California, Los Angeles, and the JPL/CalTech funded by NASA. We thank the anonymous referee for constructive suggestions which have improved the quality of the paper.

\label{lastpage}
\end{document}